\documentstyle[12pt]{article}

\begin{document}
\title{Semiclassical Gravitational Effects in the Spacetime of a Magnetic Flux 
 Cosmic String}
\author{M. E. X. Guimar\~aes  \\
\mbox{\small{Department of Physics and Astronomy }}\\
\mbox{\small{University of Wales, College of Cardiff}} \\
\mbox{\small{PO Box 913, Cardiff CF2 3YB, UK}}}
\date{}
\maketitle
\begin{abstract}
It is well-known that some physical effects may arise in the spacetime of 
a straigth cosmic string due to its global conic properties. Among these 
effects, the vacuum polarization effect has been extensively studied in the 
litterature. In papers of reference [4] a more general situation has been 
considered in which the cosmic string carries a magnetic flux $\Phi$ and 
interacts with a charged scalar field. In this case, the vacuum polarization 
arises both via non-trivial gravitational interaction 
(i.e, the conical structure) and via Aharonov-Bohm interaction. In papers [4] 
the non vanishing VEV of the energy-momentum tensor of the scalar field were 
computed. However, this energy-momentum tensor should, in principle, be taken 
into account to determine the spacetime associated with the magnetic flux 
cosmic string. Using the semiclassical approach to the Einstein eqs. we 
find the first-order (in $\hbar$) metric associated to the cosmic string 
and we show that the gravitational force resulting from the backreaction of 
the $\langle T^{\mu}_{\nu} \rangle$ is attractive 
or repulsive depending on whether the magnetic flux is 
absent or present, respectively.
\end{abstract}

In General Relativity, a 
static, straight axially symmetric cosmic string is described by the 
metric \cite{go}
\begin{equation}
ds^2 = -dt^2 +dz^2 +d\rho^2 +B^2\rho^2d\varphi^2
\end{equation}
in cylindrical coordinates $(t,z,\rho,\varphi)$ such that $\rho \geq 0$ and 
$0 \leq \varphi < 2\pi$. The constant $B$ is related to the linear mass density 
$\mu$ of the string: $B= 1-4\mu$. (We work in the system of units in which 
$G=c=1$ and $\hbar\sim 2.612 \times 10^{-66}$). For GUT strings, 
$\mu$ is of order $\mu \sim 10^{22}$ g/cm. As it is well-known, metric (1) is 
locally but not globally flat and may be written in a Minkowskian form 
with azymuthal deficit angle $\Delta = 8\pi\mu$ \cite{vi}. 
Although the cosmic string does not exert 
gravitational force on test particles, 
some physical effects may arise due solely to the global conic geometry. 
One of these effects - the vaccum polarization - has been 
extensively studied in the literature \cite{he,dow}. 
In papers of reference \cite{dow} a more general situation 
has been considered  
in which the cosmic string carries a magnetic flux $\Phi$ and interacts 
with a charged scalar field placed in the metric (1). In this case, the 
vacuum polarization effect arises not only via non-trivial gravitational 
interaction (i.e, the conical structure) but also via Aharonov-Bohm interaction.   
In papers  \cite{dow} the non-vanishing VEV of the energy-momentum tensor 
for the scalar field in the fixed background (1) were computed. However, 
this non-vanishing energy-momentum tensor should be 
taken into account to determine the spacetime metric associated with the 
magnetic flux cosmic string. This is the purpose of the present work. 
Throughout this paper we will work in the framework of the 
semiclassical approach to 
the Einstein eqs. $G_{\mu\nu}=8\pi \langle T_{\mu\nu} \rangle $ and we will 
treat this problem using the perturbative approach as in Hiscock's paper 
\cite{hi}. In this approach, the first-order (in $\hbar$) $\langle 
T_{\mu\nu} \rangle$ is treated as a matter perturbation of the zeroth-order 
metric (1) and it can be used to compute the first-order metric 
perturbation associated to it by solving the linearized Einstein's eqs. about 
the zeroth-order metric. In the present case, there will be contributions from 
both the non-trivial gravitational and the Aharonov-Bohm interactions. 
 
We start by rewriting the $\langle T^{\mu}_{\nu}\rangle$ of a massless, 
charged scalar field given in \cite{dow} as 
\begin{eqnarray}
\langle T^t_t\rangle  & = & \langle T^z_z\rangle = 
\frac{\hbar}{\rho^4}[A(\gamma)+B(\gamma)] \nonumber \\
\langle T^{\rho}_{\rho}\rangle & =  & -\frac{1}{3} 
\langle T^{\varphi}_{\varphi} \rangle =
\frac{\hbar}{\rho^4}[A(\gamma)-\frac{1}{2}B(\gamma)]
\end{eqnarray}
in terms of the dimensionless quantities
\begin{eqnarray}
A(\gamma) & \equiv  & \omega_4(\gamma) -\frac{1}{3}\omega_2(\gamma) \nonumber \\
B(\gamma)  & \equiv & 4(\xi - \frac{1}{6})\omega_2(\gamma)
\end{eqnarray}
where the constants $\omega_2(\gamma)$ and $\omega_4(\gamma)$ are defined in 
\cite{dow} (see, for instance, expressions (7.11) and (7.12) 
in Guimar\~aes and Linet) and $\gamma$ is the 
fractional part of $\{ \frac{\Phi}{\Phi_0} \}$, $\Phi_0$ is the quantum flux 
$\Phi_0 =2\pi\hbar/e$. $\gamma$ lies in the domain $0 \leq \gamma <1$, 
$\gamma = 0$ represents the case where the magnetic flux is absent. 
The $\langle T^{\mu}_{\nu}\rangle$ above is linear in $\hbar$ and its dimensionality is 
$[L]^{-2}$. We can now attempt to solve the semiclassical Einstein's 
equations $G_{\mu\nu}=8\pi\langle T_{\mu\nu}\rangle$ 
at linearized level to obtain the 
first-order metric perturbation associated to the backreaction of the 
$\langle T^{\mu}_{\nu}\rangle$ (2). We follow here the same approach as Hiscock in 
paper \cite{hi} and we set a static, 
cylindrically symmetric metric in general form
\begin{equation}
ds^2=e^{2\Phi(\rho)}(-dt^2+dz^2+d\rho^2)+e^{2\Psi(\rho)}d\varphi^2,
\end{equation}
where $\Phi$ and $\Psi$ are functions of $\rho$ only. Expanding this metric 
about the background metric we obtain the linearized Einstein eqs. with 
source (2). 
The general solutions for these equations can be easily found \cite{gui} and 
the exterior metric (corrected at first-order in $\hbar$) of the magnetic 
flux cosmic string is then obtained\footnote{We make here a change of variables 
$ r=\rho + 2\pi\frac{\hbar}{\rho}[A(\gamma)-\frac{1}{2}B(\gamma)]$ such that 
the new radial coordinate measures now the proper radius from the string.}
\begin{eqnarray}
ds^2 & = & \left[ 1-4\pi\frac{\hbar}{r^2}[A(\gamma)-\frac{1}{2}B(\gamma)] \right]
(-dt^2+dz^2)+dr^2 \nonumber \\
& & +(1-4\mu)^2r^2\left[ 1+16\pi\frac{\hbar}{r^2}
[A(\gamma)+\frac{1}{4}B(\gamma)] \right] d\varphi^2
\end{eqnarray}

The first consequence is the 
appearance of a non vanishing gravitational force on a massive test particle.
Using the definitions (3), we can obtain expressions for the gravitational force 
for both the minimal ($\xi =0$) and conformal ($\xi =1/6)$ couplings 
\begin{eqnarray*}
f^r & = & -4\pi\frac{\hbar}{r^3}\omega_4(\gamma) \\
f^r & = & -4\pi\frac{\hbar}{r^3}[\omega_4(\gamma)-\frac{1}{3}\omega_2(\gamma)] ,
\end{eqnarray*}
respectively. The first-order corrections to the deficit angle are also 
obtained. For both the minimal and conformal couplings it has the following 
expressions 
\begin{eqnarray*}
\Delta\varphi & = & 8\pi\mu -(1-4\mu)16\pi^2\frac{\hbar}{r^2}[\omega_4(\gamma)
-\frac{1}{2}\omega_2(\gamma)] \\
\Delta\varphi & = & 8\pi\mu -(1-4\mu)16\pi^2\frac{\hbar}{r^2}[\omega_4(\gamma) 
-\frac{1}{3}\omega_2(\gamma)],
\end{eqnarray*}
respectively.
 
Let us first analyse the sign of the gravitational force. 
In the case where there is no magnetic flux ($\gamma=0$) the gravitational 
force is always {\em attractice} for both minimal and conformal 
couplings. However, when the magnetic 
flux is present 
it is easy to see that the gravitational force is 
{\em repulsive} for both minimal and conformal couplings. 
Considering now the deficit angle, again the behaviour changes whereas the 
magnetic flux is present or not. When it is absent, the deficit angle 
{\em increases} ({\em decreases})  as $r\rightarrow 0$ for minimally 
(conformally) coupled scalar field. When 
the magnetic flux is present, 
the deficit angle {\em decreases} ({\em increases}) as $r\rightarrow 0$ for 
minimally (conformally) coupled scalar field.
Thus, it seems clear from these analysis that the Aharonov-Bohm interaction 
dominates over the gravitational interaction. This confirms previous 
statement by Alford and Wilczek \cite{alf}, though in slightly different context.


\begin{thebibliography}{99}


\bibitem{go} J. R. Gott III, {\em Ap. J.} {\bf 288}, 422 (1985); 
W. A. Hiscock, {\em Phys. Rev. D} {\bf 31}, 3288 (1985); B. Linet, {\em 
Gen. Rel. Grav.} {\bf 17}, 1109 (1985). 
\bibitem{vi} A. Vilenkin, {\em Phys. Rev. D} {\bf 23}, 852 (1981).
\bibitem{he} T. M. Helliwell and D. A. Konkowski, {\em Phys. Rev. D} {\bf 34}, 
1918 (1986); B. Linet, {\em Phys. Rev. D} {\bf 35}, 536 (1987); A. G. 
Smith, {\em The Formation and Evolution of Cosmic Strings}, eds. G. W. Gibbons, 
S. W. Hawking and T. Vaschaspati (Cambridge: Cambridge Univ. Press) p 680 
(1990).
\bibitem{dow}J. S. Dowker, {\em J. Phys. A} {\bf 10}, 115 (1977); 
{\em Phys. Rev. D} {\bf 36}, 3095 (1987);  
V. P. Frolov and E. M. Serebrianyi, {\em Phys. Rev. D} {\bf 35}, 
3779 (1987); 
M. E. X. Guimar\~aes and B. Linet, {\em Comm. Math. Phys.} 
{\bf 165}, 297 (1994).
\bibitem{hi}W. A. Hiscock, {\em Phys. Lett.} {\bf B188}, 317 (1987).
\bibitem{gui} M. E. X. Guimar\~aes, {\em in preparation}  (1996) .
\bibitem{alf} M. G. Alford and F. Wilczek, {\em Phys. Rev. Lett.} {\bf 62}, 1071 
(1989). 

\end{thebibliography}
\end{document}